 \documentclass[final,3p,times,twocolumn]{elsarticle}

\usepackage{array}
\usepackage{siunitx}
\usepackage{booktabs}
\usepackage{amsmath}
\usepackage{amssymb}
\usepackage{lipsum}
\usepackage{graphicx}
\graphicspath{{figure/}}
\usepackage{stfloats}
\usepackage{dcolumn}
\usepackage{bm}
\usepackage[usenames,dvipsnames]{xcolor}

\usepackage{ulem}

\usepackage[protrusion=true,expansion=true,final]{microtype}
\usepackage{xspace}
\usepackage[usenames,dvipsnames]{xcolor}
\usepackage[bookmarks=false,colorlinks]{hyperref}
\hypersetup{
    linkcolor=magenta,        
    citecolor=OliveGreen,     
    filecolor=Plum,      	
    urlcolor=Midnightblack,           
}

\biboptions{sort&compress}

\usepackage{bm}
\usepackage{soul}
\usepackage{color, xcolor}
\usepackage{multirow}
\usepackage{mathrsfs}
\begin{document}

\begin{frontmatter}



\title{Generative Inversion for Property-Targeted Materials Design: Application to Shape Memory Alloys}


\author[label1]{Cheng Li\fnref{equal}}
\address[label1]{State Key Laboratory for Mechanical Behavior of Materials, Xi'an Jiaotong University, Xi'an 710049, China}
\author[label1]{Pengfei Dang\fnref{equal}}
\author[label1]{Yuehui Xian}
\author[label1]{Yumei Zhou*}
\ead{zhouyumei@xjtu.edu.cn}
\author[label1]{Bofeng Shi}
\author[label1]{Xiangdong Ding*}
\author[label1]{Jun Sun}
\author[label1]{Dezhen Xue*}
\fntext[equal]{The two authors contributed equally to this work.}
\cortext[cor1]{Corresponding author}
\ead{xuedezhen@xjtu.edu.cn}


\begin{abstract}
The design of shape memory alloys (SMAs) with high transformation temperatures and large mechanical work output remains a longstanding challenge in functional materials engineering. 
Here, we introduce a data-driven framework based on generative adversarial network (GAN) inversion for the inverse design of high-performance SMAs. 
By coupling a pretrained GAN with a property prediction model, we perform gradient-based latent space optimization to directly generate candidate alloy compositions and processing parameters that satisfy user-defined property targets. 
The framework is experimentally validated through the synthesis and characterization of five NiTi-based SMAs. 
Among them, the Ni$_{49.8}$Ti$_{26.4}$Hf$_{18.6}$Zr$_{5.2}$ alloy achieves a high transformation temperature of 404~$^\circ$C, a large mechanical work output of 9.9~J/cm$^3$, a transformation enthalpy of 43 J/g , and a thermal hysteresis of 29 °C, outperforming existing NiTi alloys. 
The enhanced performance is attributed to a pronounced transformation volume change and a finely dispersed of Ti$_2$Ni-type precipitates, enabled by sluggish Zr and Hf diffusion, and semi-coherent interfaces with localized strain fields. 
This study demonstrates that GAN inversion offers an efficient and generalizable route for the property-targeted discovery of complex alloys.
\end{abstract}


\end{frontmatter}


\section{Introduction}

Shape memory alloys (SMAs) are functional materials that undergo reversible martensite transformations, enabling them to recover large strains and perform work output in response to mechanical stimuli~\cite{RN63,RN397}. 
They are widely employed in aerospace actuators, biomedical devices, and energy conversion systems~\cite{RN183,RN512}. 
Ultra high temperature SMAs operating reliably above 300~$^\circ$C with substantial mechanical work output are essential for mission-critical aerospace and nuclear systems, where conventional SMAs fail due to a phase transformation temperature limitation~\cite{RN178,RN350}. 
The intricate interdependencies among composition, microstructure, and processing parameters create a high-dimensional optimization challenge for achieving the targeted combination of elevated transformation temperatures, large latent heat, low thermal hysteresis, and high work output~\cite{RN178,RN184,RN396,RN579}. 

Conventional alloy design strategies, including empirical trial-and-error, thermodynamic modeling, and micromechanical simulations, have provided valuable insights but remain inherently constrained in their ability to efficiently explore high-dimensional, nonlinear design spaces.
To overcome these limitations, recent advances in data-driven materials design have employed machine learning (ML) techniques to accelerate alloy discovery and optimization~\cite{RN139,RN577}.
In the context of SMAs, several studies have developed surrogate models to predict transformation temperatures or thermal hysteresis from compositional and processing descriptors, with some predictions successfully validated through experimental synthesis~\cite{RN63,RN373,RN500,li2025knowledge,RN503,RN499,RN578}.
For example, polynomial regression models informed by Landau theory have been used to predict transformation temperatures, and Bayesian optimization has guided the design of alloys with low hysteresis, large shape memory response, and high transformation enthalpy~\cite{RN126,RN487,RN413,RN524,RN165,RN554,RN556}.
Predictive ML models have also been applied to the co-optimization of elastocaloric strength and stress hysteresis in elastocaloric alloys, demonstrating their utility for multi-objective materials design~\cite{RN395,RN147}.

Beyond forward property prediction, there is growing interest in inverse design strategies aimed at identifying input parameters that yield user-defined material properties~\cite{RN530,RN531,RN330,RN149,RN532}.
Generative models, particularly variational autoencoders (VAE)~\cite{RN428}and generative adversarial networks (GAN)~\cite{RN239}, have emerged as promising tools for this purpose in complex alloy systems, including high entropy alloys~\cite{rao2022machine,roy2023rapid,RN551,RN424}, magnesium alloys\cite{RN547}, titanium alloys~\cite{RN337,RN432}~, aluminium alloys~\cite{RN421,RN557}, metallic glasses~\cite{RN331}, dual-phase steel~\cite{RN559,RN558,RN564,RN549}, Beryllium Alloys~\cite{RN566}.
For instance, VAEs coupled with active learning have been used to identify high-entropy alloys (HEAs) with record-low thermal expansion~\cite{rao2022machine}, and GAN coupled with neural regression were used to design multi-principal-element alloys exhibiting record-high hardness~\cite{roy2023rapid}.
In the context of SMAs, a conditional VAE framework has been developed to co-design composition and processing conditions.
Notably, this approach enabled the generation of NiTi-based alloys tailored to match user-specified transformation property curves, with multiple designs successfully validated via experimental synthesis~\cite{dang2025elastocaloric}. 
Inverse design methods mentioned above employing direct condition embedding face significant risks of mode collapse when trained on limited material datasets, mirroring the challenges observed in image-based generative models~\cite{RN582,RN583}. 

In the present study, we introduce a generative inversion framework for the inverse design of high-performance SMAs based on an implicit condition fusion mechanism driven by gradient-based optimization.
Our method integrates a pretrained GAN, which captures the joint distribution of alloy compositions and processing parameters, with a property prediction model trained to estimate transformation characteristics.
By performing gradient-based optimization in the latent space of the GAN, the framework enables a direct and differentiable mapping from property space to design space.
This allows for the rapid, interpretable generation of candidate alloy designs that satisfy user-defined targets, such as transformation temperature and mechanical work output.
Although demonstrated here on SMAs, the framework is generalizable and can be extended to a broad range of material systems, including high-entropy alloys and ceramics.

\section{Design Strategy}

\subsection{Overview of the Inverse Design Framework}

To enable the inverse design of SMAs with tailored transformation temperatures and mechanical work outputs, we employ a generative inversion~\cite{RN569,RN570} framework based on a pretrained GAN. 
The key idea is to map user-specified property targets back to alloy compositions and processing parameters through gradient-based optimization in the latent space of the GAN. 

\begin{figure*}[htbp]
\begin{center}
\includegraphics[width =0.9\linewidth]{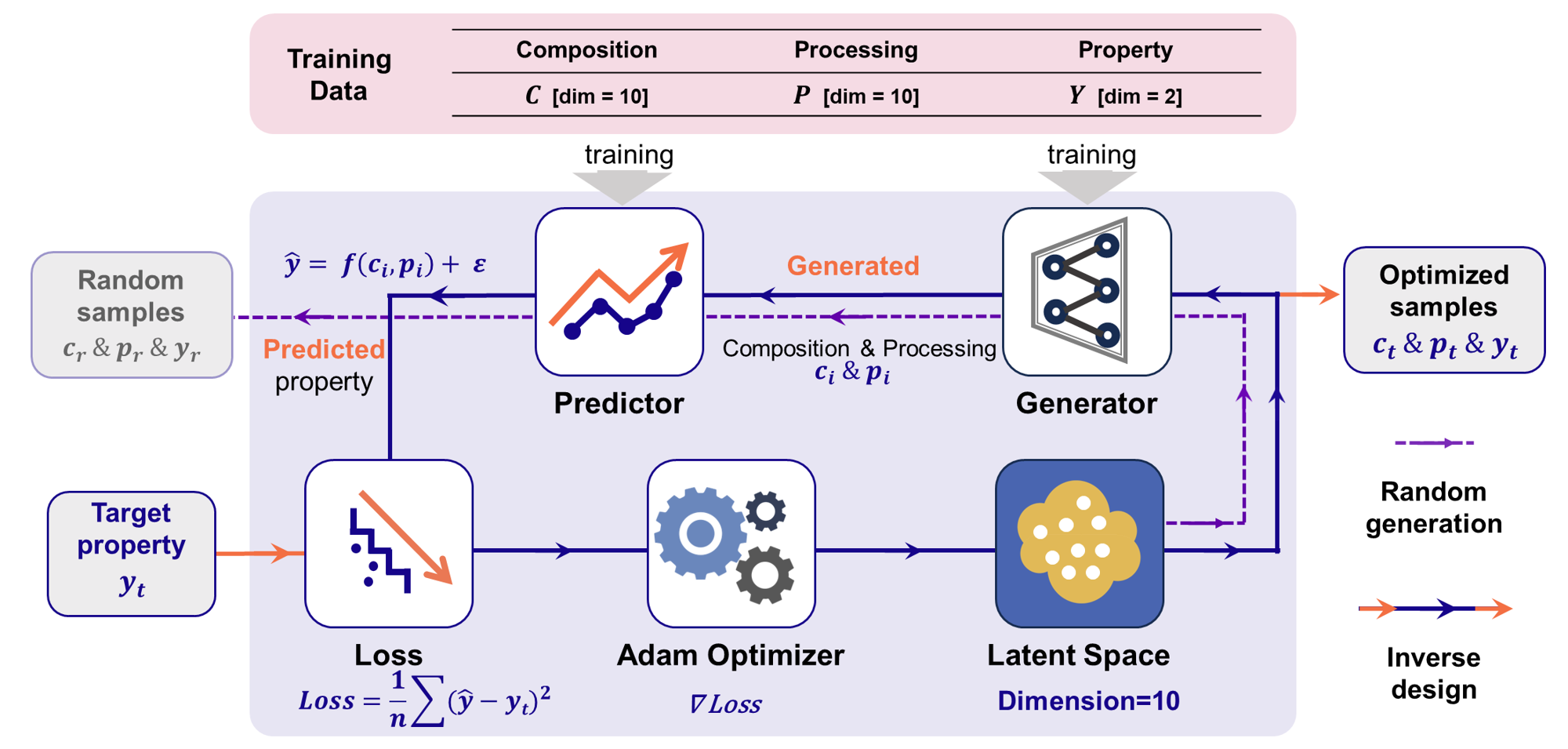}
\caption{
    \textbf{Schematic of the GAN inversion framework for property-targeted inverse design of shape memory alloys.}
    The framework integrates a pretrained GAN with a supervised property predictor to enable inverse mapping from target performance to alloy design. 
    For inverse design, a random latent vector $\mathbf{z}$ is sampled from a 10-dimensional latent space and decoded by the generator to produce candidate design vectors $(\mathbf{c}_i, \mathbf{p}_i)$. The predictor estimates the corresponding properties $\hat{\mathbf{y}} = f(\mathbf{c}_i, \mathbf{p}_i)$, and the discrepancy with the target property $\mathbf{y}_t$ is evaluated via a differentiable loss function. An Adam optimizer updates the latent code $\mathbf{z}$ to minimize the loss, guiding the generator toward design candidates whose predicted properties approach $\mathbf{y}_t$.
    The result is a set of optimized samples $(\mathbf{c}_t, \mathbf{p}_t, \hat{\mathbf{y}}_t)$ that satisfy user-defined property objectives. The same framework can also be used for unconditional random generation by sampling $\mathbf{z}$ without optimization. }
\label{fig:1}
\end{center}
\end{figure*}

This framework,  as shown in \autoref{fig:1}, comprises four main components: a generator trained to produce realistic alloy designs (i.e., composition and processing), a property predictor that estimates thermo-mechanical responses such as martensite start temperature ($M_s$) and mechanical work output, a differentiable loss function quantifies the difference between candidate samples properties and target properties, and an Adam optimizer updates the latent code $\mathbf{z}$ to minimize the loss.
By coupling generative modeling with property prediction, the framework can achieve reverse design and random generation by sampling $\mathbf{z}$ without optimization.
For inverse design, the latent vector of the GAN is iteratively optimized to minimize the loss function, producing candidates that progressively satisfy user-defined objectives. 

\vspace{0.5em}
\subsection{Latent Space Optimization for Property-Targeted Design}

{\it Generative Model and Surrogate Predictor.} We first train a Wasserstein GAN with gradient penalty (WGAN-GP)~\cite{RN568,RN570}. The generator $G\colon \mathbb{R}^d \rightarrow \mathbb{R}^n$, which maps a latent vector $\mathbf{z} \in \mathbb{R}^d$ to a design vector $\mathbf{x} = G(\mathbf{z}) \in \mathbb{R}^n$. Each design vector $\mathbf{x}$ includes both alloy composition $\mathbf{c} \in \mathbb{R}^{n_c}$ and processing parameters $\mathbf{p} \in \mathbb{R}^{n_p}$, i.e., $\mathbf{x} = (\mathbf{c}, \mathbf{p})$. The discriminator $D\colon \mathbb{R}^n \rightarrow \mathbb{R}$, outputs scalar scores for authenticity of vector $\mathbf{x}$. Specifically, ${d}$, ${n}$, ${n_c}$, and ${n_p}$  are set to 10, 19, 10, and 9, respectively. The WGAN-GP loss functions are defined as:

\begin{equation}
\mathcal{L}_D = {E}_{\tilde{x} \sim {P}_g}[D(\tilde{x})] - {E}_{x \sim {P}_r}[D(x)] + \lambda \, \mathcal{P}
\end{equation}
\begin{equation}
\mathcal P = {E}_{\hat{x} \sim {P}_{\hat{x}}}[(\|\nabla_{\hat{x}}D(\hat{x})\|_2 - 1)^2]
\end{equation}
\begin{equation}
\mathcal{L}_G = -{E}_{\tilde{x} \sim {P}_g}[D(\tilde{x})]
\end{equation}
where $P_r$ is the real data distribution, $P_g$ is the generated data distribution, and $\hat{x}$ denotes random interpolates between real and generated samples. The gradient penalty term $\mathcal{P}$ (weighted by $\lambda$=10) enforces Lipschitz continuity.
The training protocol implements asynchronous updates with distinct learning rates for the G and D, 0.0002 and 0.0008, respectively. This approach mitigates mode collapse and improves gradient stability compared to conventional GANs.

In parallel, a artificial neural network (ANN) surrogate model $f\colon \mathbb{R}^n \rightarrow \mathbb{R}^m$ is trained to predict material properties $\mathbf{y} \in \mathbb{R}^m$, such as $M_s$ and mechanical work output (the product of stress and recoverable phase transformation strain). The predictor loss functions are defined as:
\begin{equation}
\mathcal{L}_{\text{surrogate}} = \frac{1}{N} \sum_{i=1}^{N} \| f(\mathbf{x}_i) - \mathbf{y}_i \|^2 +\beta \mathcal{L}_{\text{knowledge}},
\end{equation}

\noindent where $(\mathbf{x}_i, \mathbf{y}_i)$ are training pairs, ${L}_{\text{knowledge}}$ is domain-informed constraint derived from our prior studies~\cite{li2025knowledge}, $\beta$ is weighting coefficient for the domain-informed constraint, set to 5 in this work. This loss function maintains data fidelity while enforcing physical consistency. 
The WGAN-GP and predictor are trained using a dataset of known composition–processing-properties pairs, totaling 750 data points, derived from experimental measurements in our laboratory and literature data~\cite{RN283,RN73,RN299,RN273,RN266,RN275,RN277,RN301,RN265,RN276,RN279,RN284,RN269,RN272,RN406,RN278,RN270,RN282,RN268,RN9,RN13,RN16,RN23,RN38,RN261,RN262,RN48,RN210,RN274,RN158,RN157,RN350}. The parameter settings for WGAN-GP and surrogate models are detailed in the Supplementary Information \textcolor{blue}{Table~S1 and S2}.

\vspace{0.5em}
{\it Inverse Design via Latent Space Optimization.} Given a target property vector $\mathbf{y}_t \in \mathbb{R}^m$, we aim to find a latent code $\mathbf{z}^*$ such that the generated design $\mathbf{x}^* = G(\mathbf{z}^*)$ yields predicted properties $\hat{\mathbf{y}} = f(\mathbf{x}^*)$ that match the target. This leads to the optimization problem:

\begin{equation}
\mathbf{z}^* = \arg\min_{\mathbf{z}} \mathcal{L}(\mathbf{z}) = \arg\min_{\mathbf{z}} \| f(G(\mathbf{z})) - \mathbf{y}_t \|^2.
\end{equation}

\vspace{0.5em}
{\it Optimization Procedure.} This loss is minimized using gradient-based methods such as the Adam optimizer, with updates of the form:

\begin{equation}
\mathbf{z}_{k+1} = \mathbf{z}_k - \eta \cdot \nabla_{\mathbf{z}} \mathcal{L}_(\mathbf{z}_k),
\end{equation}

\noindent where $\eta$ is the learning rate, set to 0.04 in this work. Optimization is initialized with a random latent vector $\mathbf{z}_0 \sim \mathcal{N}(0, I)$ and proceeds until the predicted property satisfies a convergence criterion:

\begin{equation}
\| f(G(\mathbf{z}^*)) - \mathbf{y}_t \| < \epsilon.
\end{equation}

\vspace{0.5em}
{\it Final Design Selection.} Once converged, the optimized latent vector $\mathbf{z}^*$ is decoded to obtain the final alloy design:

\begin{equation}
\mathbf{x}^* = G(\mathbf{z}^*) = (\mathbf{c}^*, \mathbf{p}^*), \quad \hat{\mathbf{y}}^* = f(\mathbf{x}^*) \approx \mathbf{y}_t.
\end{equation}

\noindent The final design $\mathbf{x}^*$ can then be validated through high-fidelity simulations or experimental synthesis to confirm its thermo-mechanical performance.
A detailed pseudocode outlining the GAN inversion steps is provided in the Supplementary Information (\textcolor{blue}{Table~S3}). This approach enables efficient, interpretable, and experimentally accessible alloy design across coupled compositional, processing, and property spaces.

\subsection{Experimental}

New alloys were synthesized according to the chemically optimized compositions derived from our design strategy. High-purity elements (99.9 at.\% Ti, Ni, and Cu; 99.5 at.\% Hf and Zr) along with a Ti-60Ta wt.\%master alloy (XI'AN Rare Metal Materials Institute Co. Ltd) were arc-melted under argon atmosphere using a water-cooled copper crucible. To ensure compositional homogeneity, each ingot was remelted six times followed by flipping between melts.
These alloys were then heat treated according to the recommended temperature, time, and cooling method of the GAN inversion. 

Transformation characteristics were evaluated using differential scanning calorimetry (DSC; Netzsch 214) with heating/cooling rates of 10 °C/min.
Shape memory properties were characterized through uniaxial constant-stress thermal cycling tests performed on dog-bone specimens using a uTS tensile testing system (10~°C/min). Temperature monitoring was achieved via a welded K-type thermocouple, while strain evolution was tracked using digital image correlation (DIC) with a stochastic white speckle pattern applied to the gauge section.

Phase analysis was conducted via transmission wide-angle X-ray diffraction (XRD; NANOPIX-WE system, Rigaku Co., Tokyo, Japan) employing a Mo rotating anode source ($\lambda$=0.7093~\AA). Microstructural characterization was performed using a scanning electron microscopy (SEM; Zeiss sigma360) and a transmission electron microscopy (TEM; JEM F200) at 200 kV. TEM specimens were prepared by twin-jet electropolishing in a 10~vol\% $\rm HClO_4$ and 90~vol\% $\rm CH_3OH$ electrolyte maintained at \text{-}20 ~°C.

\section{Results and Discussion}
\subsection{Statistical Reconstruction Fidelity and Diversity Metrics of Generator}

We first assess the capacity of pretrained GAN to replicate the complex, high-dimensional relationships between compositions, processing parameters, and their properties, as this represents a critical prerequisite for effective inverse design applications.
We use diversity and falsity metrics to determine the optimal parameter state of the pretrained GAN, as shown in \textcolor{blue}{Supplementary Information (Figure~S1)}.

\begin{figure*}[htbp]
\begin{center}
\includegraphics[width =0.9\linewidth]{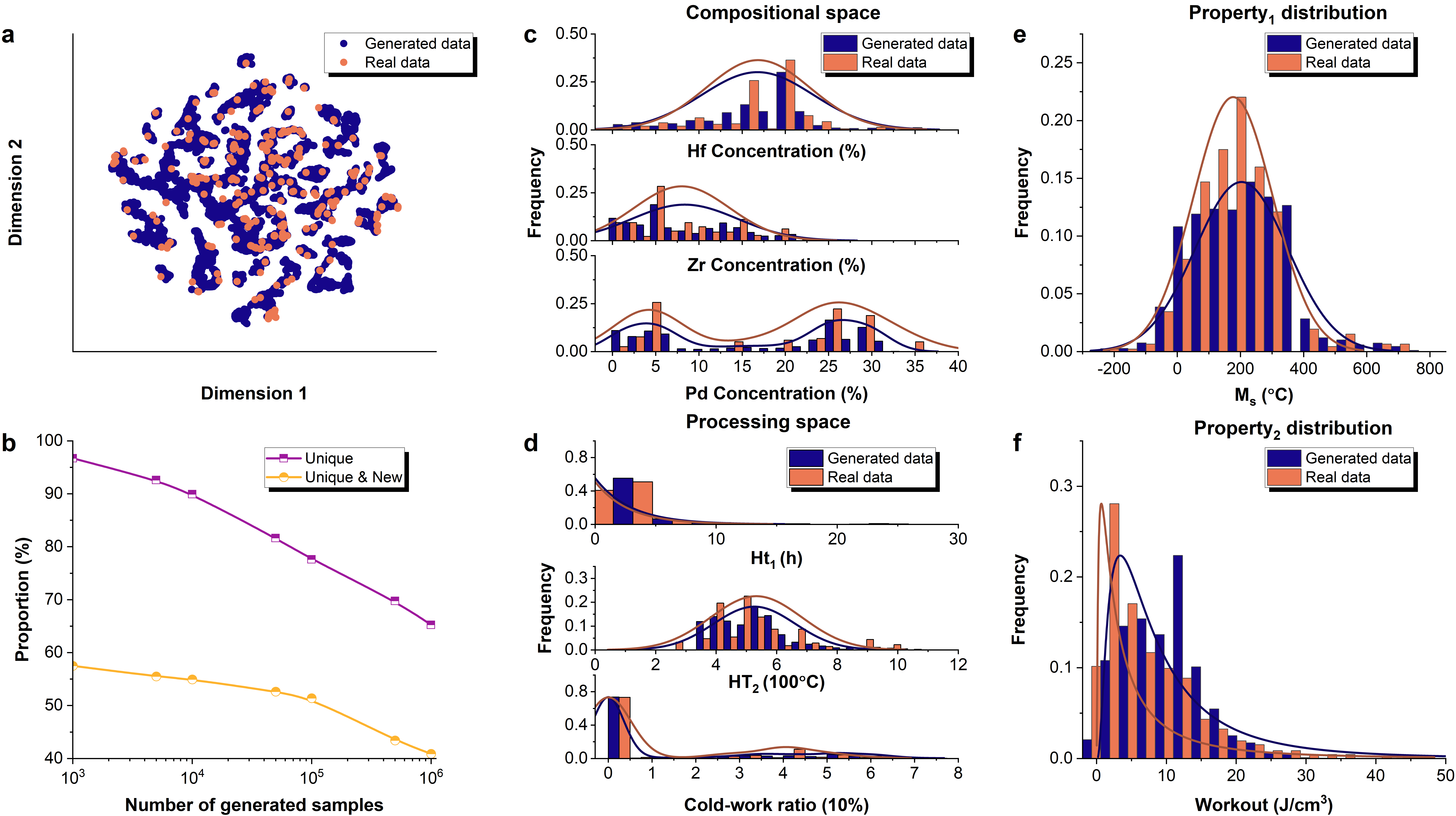}
\caption{
\textbf{Evaluation of the generative performance of the trained GAN model.}
(a) t-SNE visualization of real (orange) and GAN-generated (blue) samples in compositional and processing parameters space, showing strong distributional overlap and indicating that the generator captures the underlying data manifold.
(b) Diversity analysis of generated samples as a function of generation size. ``Unique" and ``Unique \& New'' score reflects the model’s ability to generate novel designs.
Comparison of marginal distributions for key alloying elements (Hf, Zr, Pd) (c) , processing parameters (primary heat treatment time, secondary heat treatment temperature, and cold-work ratio) (d), key material properties ($M_s$ and workout) (e)--(f) between generated and real samples, demonstrating preservation of data features. 
}
\label{fig:2}
\end{center}
\end{figure*}

To assess structural fidelity, we project both real and generated samples into a shared two-dimensional embedding using t-SNE (\autoref{fig:2}a). 
Real samples are shown in orange and GAN-generated samples in blue. 
The high degree of overlap between the two distributions indicates that the generator learns a meaningful approximation of the underlying design manifold, preserving global and local structure in the composition–processing space.

We then assess the Unique and Unique \& New metric of the generated samples (Figure~\ref{fig:2}b) based on cosine similarity.  "Unique" refers to non-redundant samples and "Unique \& New" are samples unique and not present in the training data.
Although the fraction of both unique and unique \& new shows moderate decay on scale, the generator maintains robust novelty retention, yielding 40~\% unique and previously unseen samples even at ${10 ^ 6}$ samples. This behavior reflects fundamental probability constraints, as sample volume increases, the likelihood of sampling similar latent space points rises proportionally, particularly in high-density distribution regions, consistent with state-of-the-art generative models~\cite{RN530}. 

Next, we examine whether the model reproduces the statistical characteristics of key design variables. 
\autoref{fig:2}c compares the marginal distributions of representative alloying elements (Hf, Zr, and Pd) between real and generated data. 
The generator maintains the multimodal character and relative frequency of each composition feature, indicating a strong match to the empirical compositional priors. 
Similarly, \autoref{fig:2}d shows that processing parameters such as primary heat treatment time, secondary heat treatment temperature, and cold-work ratio are faithfully reproduced. 

Finally, we evaluate whether the generator can produce realistic property values based on surrogate property predictor, which maps design vectors to thermo-mechanical responses. 
On the held-out test set, the predictor achieves a coefficient of determination of $R^2$ of 0.90 and a Root Mean Square Error (RMSE) of 43.89~$^\circ$C for $M_s$, and $R^2$ of 0.84 with an RMSE of 2.13~J/cm\textsuperscript{3} for mechanical work output.
These results confirm the surrogate’s ability to learn the underlying elements-processing and property relationships with sufficient precision for guiding latent space optimization. Detailed error analyses are provided in the \textcolor{blue}{Supplementary Information (Figure~S2–S4)}.
\autoref{fig:2}e--f compare the distributions of $M_s$ and mechanical work output between generated and real samples. 
The generated data spans the full range of observed property values, from low to high $M_s$ and from low to high actuation workout. Other distributions of elements and processing parameters between real and generated data are provided in the \textcolor{blue}{Supplementary Information (Figure~S5 and S6).}
The above analysis reflects a desirable balance: the generator maintains high fidelity to the training data while also enabling exploration of novel designs beyond the observed dataset.

\subsection{Latent Space Optimization Toward Targeted Thermo-Mechanical Properties}

To evaluate capability of the proposed framework for inverse design, we examine two representative inverse design trajectories generated by GAN inversion: one targeting elevated $M_s$, and the other targeting increased mechanical work output. 
\begin{figure}[htbp]
\begin{center}
\includegraphics[width = 1\columnwidth]{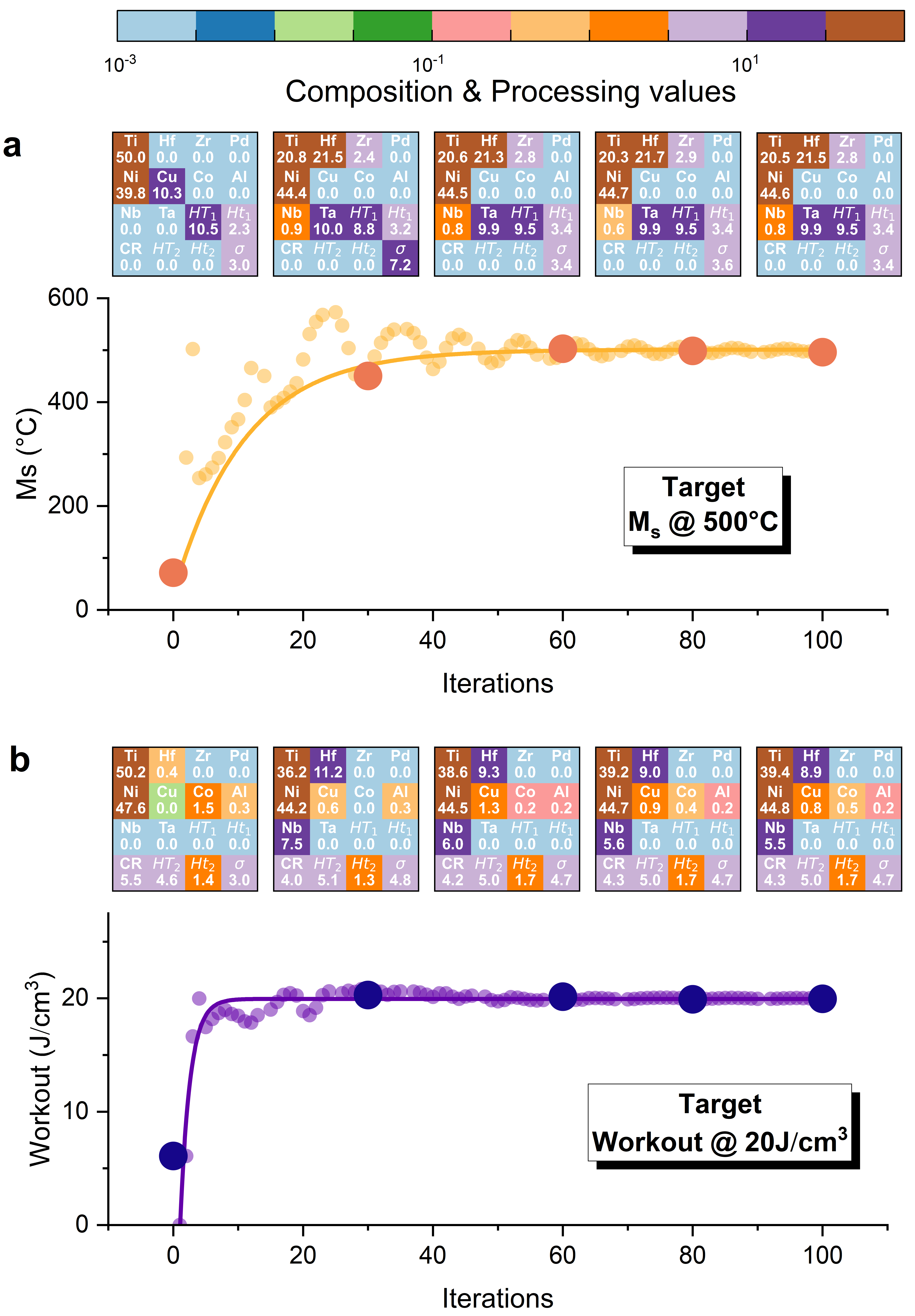}
\caption{
\textbf{Latent space optimization trajectories for property-targeted inverse design using GAN inversion.} 
(a) Optimization of $M_s$ starting from a randomly initialized latent vector. Each point represents the predicted $M_s$ from the surrogate model at a given iteration. The inversion process progressively shifts the design toward alloys predicted to exceed $500\,^\circ$C in $M_s$.
(b) Optimization targeting increased mechanical work output. As in panel (a), the inversion iteratively updates the latent vector to maximize predicted workout, approaching the specified target of $20$~J/cm\textsuperscript{3}. 
The curve traces the optimization path, and representative design snapshots reveal how compositional and processing features evolve to support enhanced performance.
}
\label{fig:3}

\end{center}
\end{figure}
\autoref{fig:3}a illustrates the optimization of $M_s$ beginning from a randomly sampled latent vector $\mathbf{z}_0$. 
At each iteration, the generator produces a candidate design $\mathbf{x}_i = G(\mathbf{z}_i)$, which is evaluated by the surrogate model to predict the corresponding $M_s$. 
The curve shows a smooth, monotonic increase in predicted $M_s$, ultimately converging toward the target value of $500\,^\circ$C. 
This behavior highlights the differentiability and directionality of the latent space, enabling efficient navigation toward high-temperature transformation regimes.

To gain insight into the underlying design evolution, five selected designs along the trajectory are visualized in terms of their compositional and processing attributes by heatmap. 
 The heatmap values correspond to atomic percentages (for compositions) and scaled units (for processing variables), with full variable definitions provided in \textcolor{blue}{the Supplementary Information (Table~S4)}.
These heatmaps reveal systematic trends, such as increased Hf, Zr and Ta content or adjustments in heat treatment parameters, that correlate with rising $M_s$. 
\autoref{fig:3}b shows a parallel optimization targeting mechanical workout. 
Starting again from a random initialization, the inversion trajectory drives the predicted workout toward a target of 20~J/cm\textsuperscript{3}, with rapid improvement observed in the early stages. 
As in the temperature-targeting example, the evolving design vectors are visualized as composition–processing heatmaps. 
The optimization induces increases in elements and processing conditions known to support high actuation stress and strain recovery.

The observed trajectories demonstrate robust convergence behavior and physically interpretable design pathways, validating the capability of framework capability for targeted property optimization in shape memory alloys. 
These results establish the efficacy of framework in addressing inverse design challenges across complex, nonlinear, and multi-variable materials systems.

\subsection{Targeted Property Control and Design Diversity through GAN Inversion}

Here, we evaluate the ability of  the framework to generate designs that are both precise, novel, and diverse in chemistry and processing space.

\begin{figure*}[htbp]
\begin{center}
\includegraphics[width =0.8\linewidth]{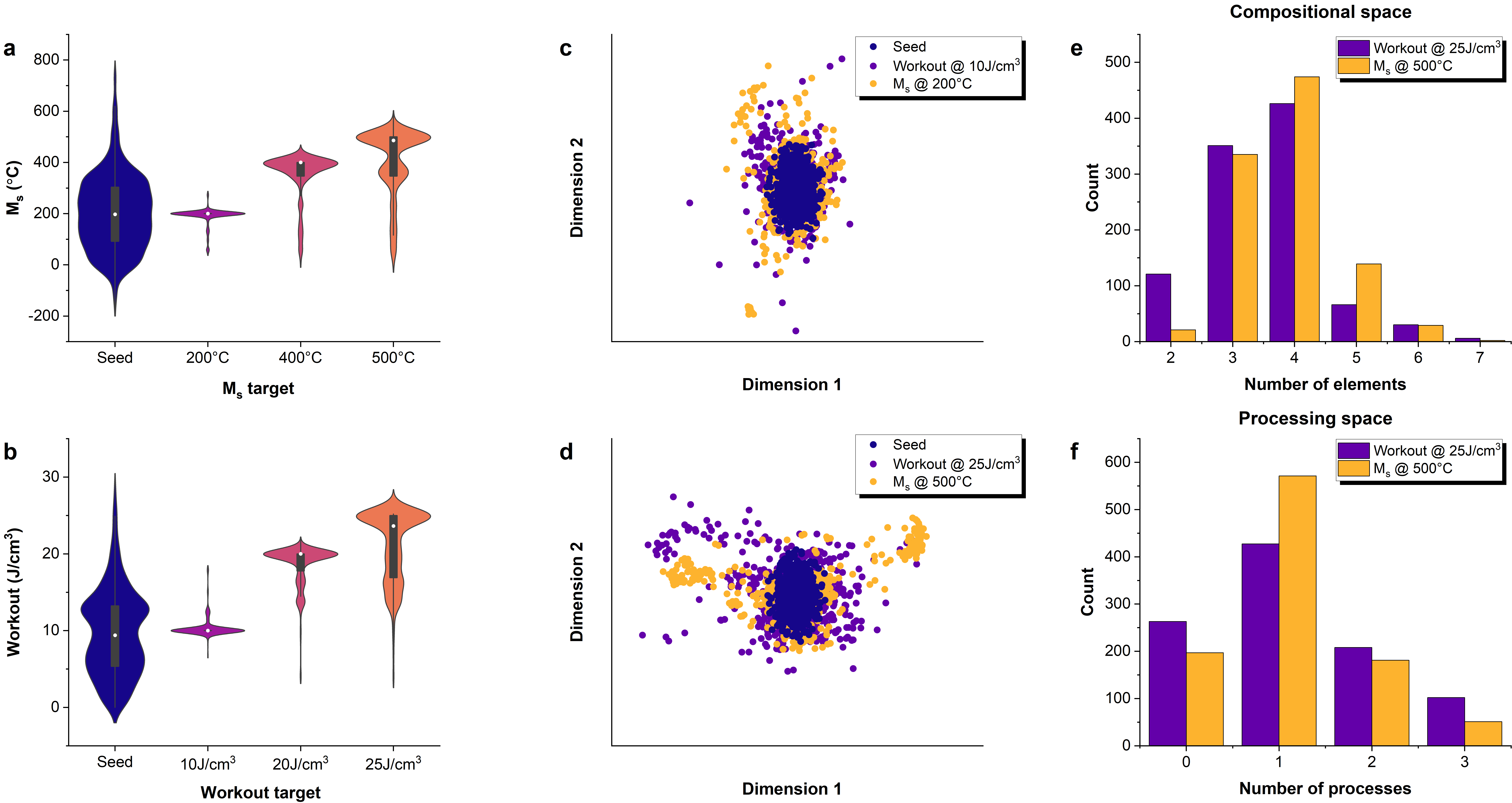}
\caption{
\textbf{Property-targeted conditional generation: precise, novel, and diverse designs produced via GAN inversion.}
{\bf (a–b).} Violin plots of generated samples targeting 
(a) martensite start temperature ($M_s$ = 200, 400, 500 $^\circ$C) and (b) mechanical work output (10, 20, 25 J/cm$^3$). The narrowing of distributions and alignment of medians with target values demonstrate precise property control, in contrast to the broad baseline distributions of the initial seeds. 
{\bf (c–d).} Principal component analysis (PCA) projections comparing seed samples with inversion-generated designs for distinct objectives.
(c) Designs targeting $M_s$ = 200~$^\circ$C and workout = 10~J/cm$^3$ show clear directional shifts in design space. 
(d) Designs optimized for more extreme targets ($M_s$ = 500~$^\circ$C, workout = 25~J/cm$^3$) diverge further from the seed distribution, indicating exploration of novel regions.
{\bf (e–f).} Histograms showing the distribution of (e) number of elements in alloy composition and (f) number of processing steps for two high-performance targets: $M_s$ = 500~$^\circ$C (gold) and workout = 25~J/cm$^3$ (violet). The distinct distributions show diversity of designs in both chemical and processing space.
}
\label{fig:4}
\end{center}
\end{figure*}

To assess precision, we perform conditional generation for a range of martensite start temperature ($M_s$) and mechanical work output targets. 
As shown in \autoref{fig:4}a--b, the predicted property distributions of generated samples are tightly centered around the specified targets, clearly shifted away from the broad baseline of the seed distribution. 
For example, $M_s$ values generated for targets of 200, 400, and 500~$^\circ$C exhibit increasing medians that align with their respective objectives. 
However, when targeting more extreme values (such as 500~$^\circ$C), the variance becomes larger, reflecting reduced precision caused by limited training data in the high-temperature regime. 
Similar behavior is observed in the workout-targeted samples. 
While the medians consistently reach their objectives, the spread increases at higher target values.

To explore the latent dynamics of conditional generation, we apply principal component analysis (PCA) to the design vectors (\autoref{fig:4}c--d). 
Compared to the dense cluster of initial seed points (navy), the generated samples exhibit distinct directional shifts depending on the targeted property. 
For instance, designs optimized for high $M_s$  and high workout  occupy different regions of the PCA space. 
These shifts indicate that the model learns to explore separable areas of the design manifold that correspond to different thermo-mechanical objectives, reflecting distinct chemical and processing strategies.

We further assess design diversity using histograms of alloying element count and processing complexity (\autoref{fig:4}e--f). 
Alloys optimized for high mechanical work output tend to involve more elaborate processing protocols, often incorporating two or more distinct heat treatments.
In contrast, high $M_s$ designs are more likely to rely on simpler procedures such as single-step thermal cycles. 
On the compositional side, both target classes cluster around three to five elements, but their distributions differ in their tails. 
This suggests that different property objectives favor different degrees of chemical complexity.

Together, these results confirm that GAN inversion supports both accurate property targeting and exploration of meaningful design strategies. 
%

\subsection{Experimental Validation of GAN inversion Designed Alloys}

\begin{figure*}[htbp]
\begin{center}
\includegraphics[width =0.8\linewidth]{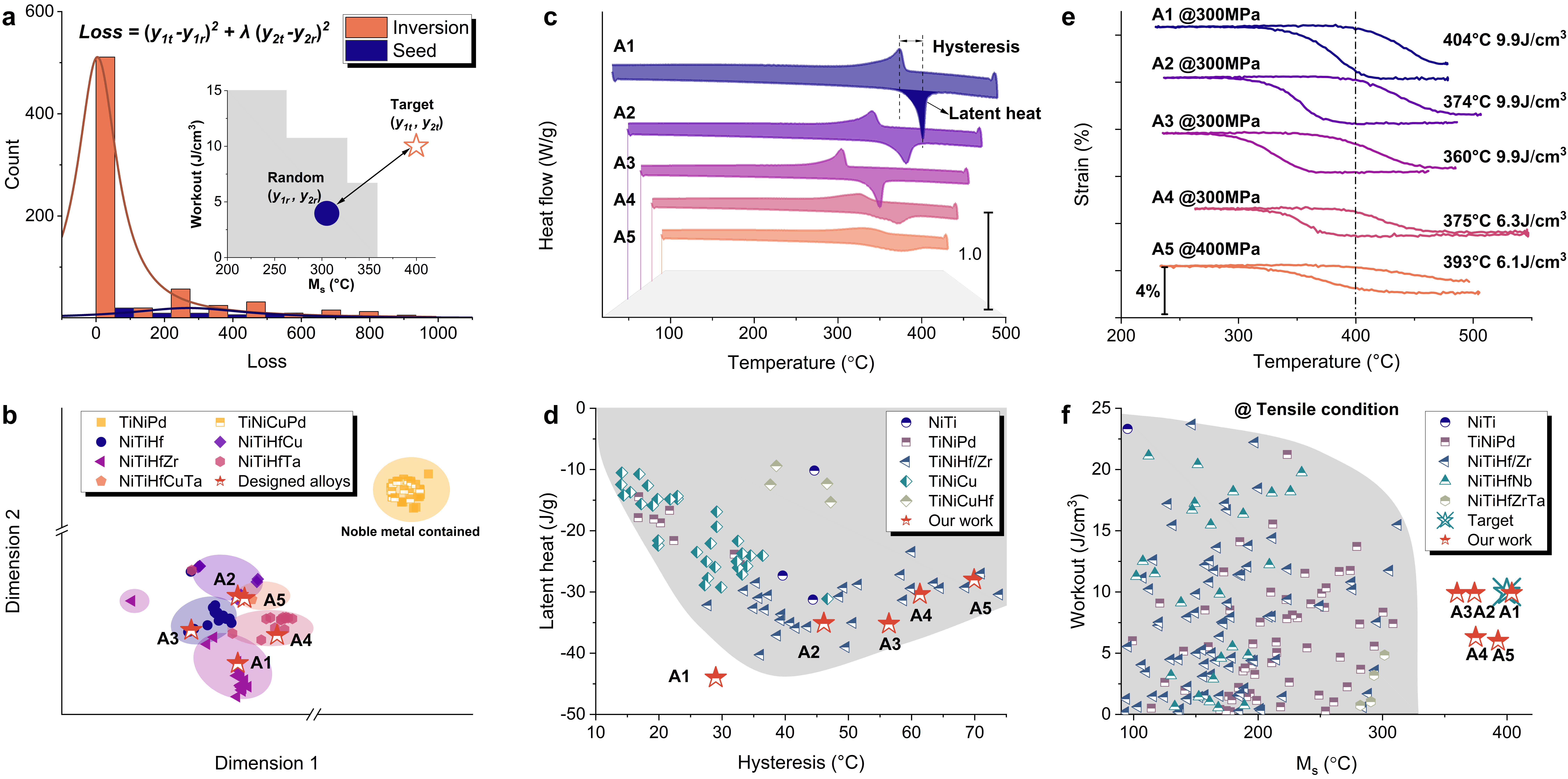}
\caption{
\textbf{Experimental validation of GAN inversion-designed high-temperature shape memory alloys.} 
(a) Distribution of loss, defined as the mean squared error between predicted and target property values, for randomly generated seed samples versus GAN-inversion-optimized candidates. Inset shows conceptual depiction of the inversion objective in the $M_s$–workout space . 
(b) t-SNE projection of alloys designed by GAN inversion that met the target in composition-processing space. Colors denote alloy families, and stars highlight the experimental validation samples. 
(c) DSC profiles of the five synthesized alloys (A1–A5) . 
(d) Performance comparison of A1–A5 alloys against literature systems in the latent heat versus hysteresis space. 
(e) Thermomechanical characterization under applied tensile stress (300–400~MPa) between 250~$^\circ$C and~510~$^\circ$C. 
(f) Performance comparison of A1–A5 alloys against target property and literature systems mechanical workout versus $M_s$ under tensile stress conditions. 
}
\label{fig:5}
\end{center}
\end{figure*}

We further adopt a dual target design combined with experimental verification to verify the effectiveness of GAN inversion. 
Our target is set at $M_s$ equal to 400~$^\circ$C and workout equal to~10 J/cm$^3$, it is an advanced level of high-temperature shape memory alloys.
We first evaluate the effectiveness of the GAN inversion in reducing design error . 
As shown in \autoref{fig:5}a, the distribution of prediction loss, quantified as the mean squared error between predicted and target properties, is sharply and tall skewed toward lower values for GAN-optimized designs (orange) compared to randomly sampled seeds (navy). 
This demonstrates the framework’s ability to efficiently locate high-quality candidates within the high-dimensional latent space. 
The inset schematic illustrates the minimization objective in the $M_s$–workout plane.

\begin{table*}[t]
    \centering
    \footnotesize
    \caption{Composition, processing, and properties details of alloys A1–A5.}
    \begin{tabular}{cccccccc}
    \toprule
\textbf{ID} & \textbf{Composition} & \textbf{Processing} & \textbf{Latent heat} & \textbf{Hysteresis} & \textbf{Stress} & \textbf{M\textsubscript{s}} & \textbf{Workout} \\
& & & \textbf{(J/g)} & \textbf{($^\circ$C)}& \textbf{(MPa)} & \textbf{($^\circ$C)} & \textbf{(J/cm\textsuperscript{3})} \\
\midrule
    A1 & Ni\textsubscript{49.8}Ti\textsubscript{26.4}Hf\textsubscript{18.6}Zr\textsubscript{5.2} & 1000~$^\circ$C~2.3~h~WQ& 43    & 29.0 & 300 & 404 & 9.9 \\
    A2 & Ni\textsubscript{46.5}Ti\textsubscript{25.1}Hf\textsubscript{25.2}Cu\textsubscript{3.1} & 830~$^\circ$C 2.3\,h~AC& 35.1  & 46.1 & 300 & 374 & 9.9 \\
    A3 & Ni\textsubscript{50.1}Ti\textsubscript{26.6}Hf\textsubscript{23.3}       & 630~$^\circ$C 9.8\,h~WQ& 35.2  & 56.4 & 300 & 360 & 9.9 \\
    A4 & Ni\textsubscript{45.5}Ti\textsubscript{25.7}Hf\textsubscript{24.2}Cu\textsubscript{3.2}Ta\textsubscript{1.4} & 920~$^\circ$C 2.7\,h~WQ& 30.4 & 61.3 & 300 & 375 & 6.3 \\
    A5 & Ni\textsubscript{45.8}Ti\textsubscript{25.7}Hf\textsubscript{19.1}Zr\textsubscript{2.7}Ta\textsubscript{6.7} & 850\,$^\circ$C 2.4\,h~WQ& 28    & 69.9 & 400 & 393 & 6.1 \\
\bottomrule
    \end{tabular}
\label{table:1}
\end{table*}

Following removal of duplicate candidates from alloys designed by GAN inversion that met the target, 97 distinct alloy candidates were retained (\textcolor{blue}{Supplementary Information Table~S5}).
Dimensionality reduction analysis using t-SNE (\autoref{fig:5}b) revealed the design space encompasses seven alloys systems: Ti-Ni-Pd, Ti-Ni-Cu-Pd, Ni-Ti-Hf, Ni-Ti-Hf-Cu, Ni-Ti-Hf-Zr, Ni-Ti-Hf-Ta, and Ni-Ti-Hf-Cu-Ta. 
This outcome supports the framework’s capacity to generate chemically diverse and non-trivial designs. 
Considering the economic implications of precious metal constituents (Pd) in Ti-Ni-Pd and Ti-Ni-Cu-Pd systems, experimental validation was conducted on representative alloys selected from the remaining five systems. 

Thermal properties of the synthesized alloys were characterized using differential scanning calorimetry (DSC), as shown in \autoref{fig:5}c. 
All five samples exhibit clear $M_s$ ranging from 322 to 399~$^\circ$C, latent heat ranging from -28 to -43~J/g, and thermal hysteresis ranging from 29 to 69.9~$^\circ$C under stress free conditions. 
These results indicate tunable transformation energetics enabled by alloy design.
To benchmark thermodynamic efficiency, \autoref{fig:5}d maps the experimental alloys in the latent heat versus hysteresis domain, alongside literature data, including NiTi~\cite{RN575}, TiNiPd~\cite{RN22}, TiNiHf/Zr~\cite{RN373,RN413,RN22}, TiNiCu~\cite{RN22}, TiNiCuHf~\cite{li2025knowledge,RN576}.
The GAN inversion designed alloys occupy favorable regions near the performance frontier, the alloy A1 exhibits a latent heat of 43~J/g and a thermal hysteresis of 29~°C, outperforming existing NiTi alloys, achieving a rare combination of high latent heat and reduced hysteresis. 

Mechanical performance was evaluated by thermal cycling under tensile stress, as shown in \autoref{fig:5}e.
Alloys A1–A5 demonstrate reversible shape recovery with $M_s$ ranging from 362 to 404~$^\circ$C and mechanical workout ranging from 6.1 and 9.9 J/cm$^3$ under 300 MPa or 400 MPa tensile stress. Alloy A1 achieved the design target with $M_s$ of 404 ℃ and workout of 10J/cm$^3$.
These values place the synthesized alloys among the most effective high-temperature SMA systems reported to date.

Finally, \autoref{fig:5}f situates the experimental alloys within the $M_s$ and workout landscape along with known systems under tensile stress, including NiTi\cite{RN266,RN265}, TiNiPd\cite{RN274,RN275,RN276,RN279}, NiTiHf/Zr \cite{RN13,RN16,RN38,RN283,RN73,RN299,RN273,RN277,RN301,RN261,RN284}, NiTiHfNb\cite{RN9}, NiTiHfZrTa\cite{RN284}. The complete composition, processing, and properties details of alloys A1–A5 are provided in \autoref{table:1}.
The GAN inversion designed materials populate a previously inaccessible regime characterized by the simultaneous achievement of elevated transformation temperatures and high mechanical work densities. 
This region is highly desirable for actuation applications in extreme environments.

Together, these results confirm that the GAN inversion framework can translate computational design outputs into experimentally validated, high-performance SMAs. 

\begin{figure*}[htbp]
\begin{center}
\includegraphics[width =0.9\linewidth]{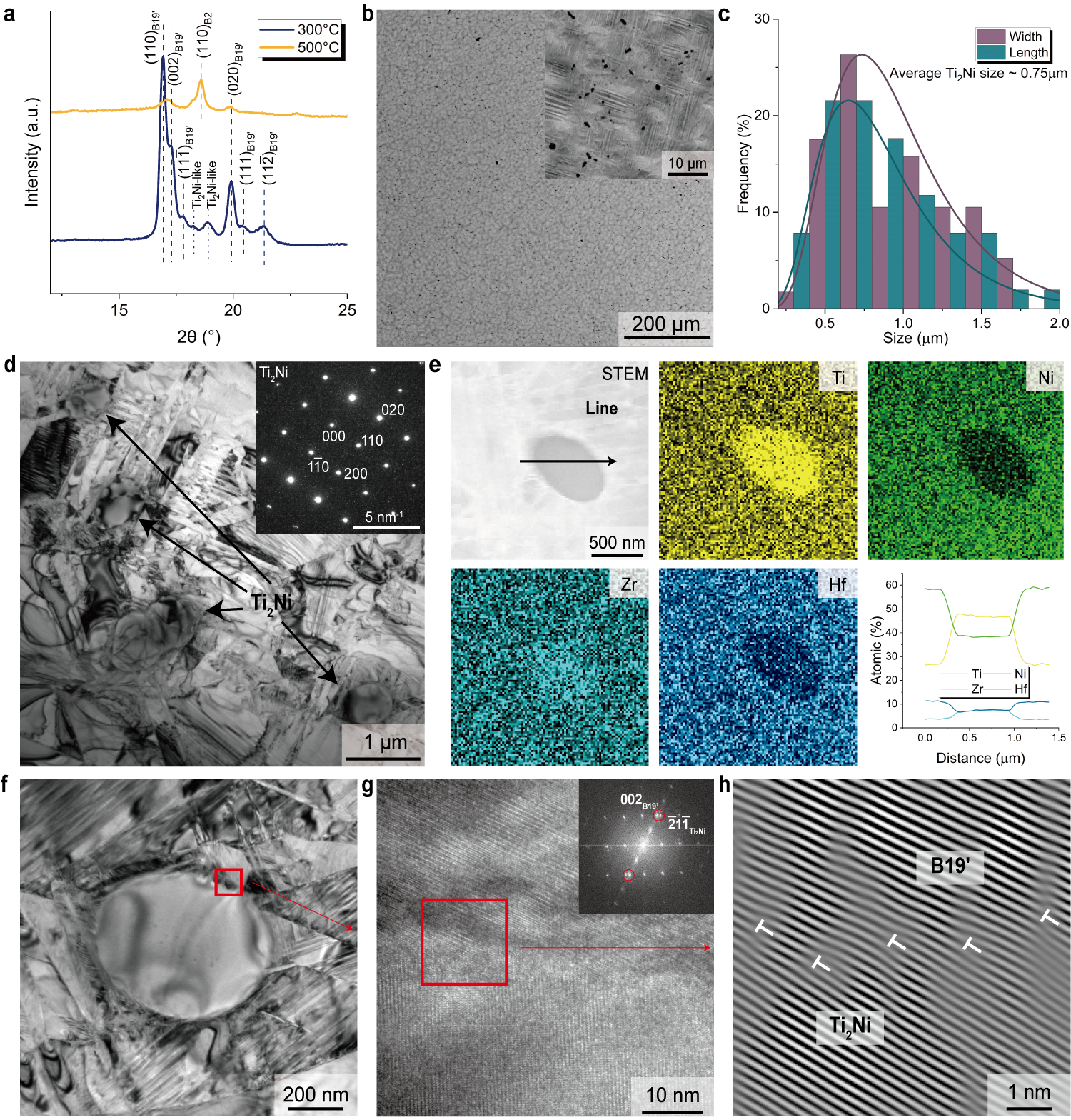}
\caption{
\textbf{Crystallographic and microstructural origin of high performance in designed Alloy A1.}
(a) XRD patterns collected at 300\,$^\circ$C and 500\,$^\circ$C, indicate the presence of B19$'$ martensite phase at 300\,$^\circ$C  and B2 Austenite phase at 500~$^\circ$C. 
(b) BSE image of the alloy, showing a homogenous dispersion of fine precipitates across the matrix. Inset shows a higher-magnification view.
(c) Histogram of precipitate size distribution, measured along both width and length directions.
(d) Bright-field TEM image shows Ti$_2$Ni-type precipitates in the matrix. Inset shows the corresponding SAED pattern of Ti2Ni-type precipitate via [001] zone axis.
(e) EDS elemental analysis of the Ti2Ni-type precipitate. 
(f) The magnified bright field image near a Ti$_2$Ni-type precipitate. (g) HR-TEM of the precipitate–matrix interface. Inset shows the corresponding FFT pattern.
(h) IFFT image of the red dashed area in (g) revealing local dislocations at the precipitate–matrix semi-coherent interface. 
}
\label{fig:6}
\end{center}
\end{figure*}

\subsection{Crystallographic and Microstructural Basis for High Performance in Alloy A1}

To elucidate the mechanisms underlying the exceptional actuation performance of Alloy A1, we conducted a detailed structural and microstructural analysis. 
The X-ray diffraction (XRD) patterns acquired at 300\,$^\circ$C and 500\,$^\circ$C are presented in \autoref{fig:6}a. The diffraction data yielded the following lattice parameters: $a_0$ = 3.1050~Å for the B2 parent phase, and $a$ = 3.046 \AA, $b$ = 4.099 \AA, $c$ = 4.825 \AA , $\beta$ = 102.17~$^\circ$ for the B19$'$ martensite phase. 
The transformation volume change was determined to be 1.66~\%, which is significantly larger than those reported for $\rm Ni_{50.3}Ti_{35}Hf_{20}$ and $\rm Ti_{50.1}Ni_{49.9}$ alloys~\cite{shuitcev2020volume}. This enhanced volume effect during phase transformation contributes to greater transformation enthalpy and high actuation strain under elevated temperatures. 

Back-scattering scanning electron (BSE) microscopy image in \autoref{fig:6}b reveals a uniform distribution of fine precipitates embedded within the matrix. 
%
The size distribution, quantified in \autoref{fig:6}c, shows a narrow range centered around 0.75 $\mu$m. 
Further microstructural characterization through transmission electron microscopy (TEM) combined with selected area electron diffraction (SAED) in \autoref{fig:6}d identifies the fine precipitates as cubic Ti$_2$Ni-type phases.
Energy-dispersive X-ray spectroscopy (EDS) and scanning TEM (STEM) mapping in \autoref{fig:6}e shows that the precipitates are compositionally significant enrichment of Ti and Zr, whereas Ni and Hf exhibit depletion. The accompanying line profile further quantitatively illustrates elemental segregation across the precipitate. 
%
Their relatively small size is attributed to the kinetic limitation of the diffusion of Zr and Hf during solidification due to their large atomic radius~\cite{RN1,RN23,RN29}. 
As the precipitates form, Hf is expelled into the surrounding matrix and Zr is extracted from the surrounding matrix; however, due to its sluggish diffusivity, Zr and Hf remains localized, limiting precipitate coarsening. 
These fine precipitates act as barriers to dislocation motion and suppress plastic deformation, thereby improving transformation reversibility under high biased stress. 

\autoref{fig:6}g shows a high-resolution TEM (HRTEM) image of the interface between Ti$_2$Ni and B19$'$ matrix \autoref{fig:6}f. The corresponding fast Fourier transformation (FFT) pattern demonstrates a semi-coherent interface, evidenced by the near-overlapping  (002)$\rm _{B19'}$ and $(\bar{2}1\bar{1})\rm_{Ti_2Ni}$ reflections. 
The inverse FFT (IFFT) image of the interface region (\autoref{fig:6}h) displays strained boundaries where interfacial dislocations are located.
These dislocations generate localized stress fields that can lower energy barriers for transformation, thus supporting transformation reversibility and reducing thermal hysteresis~\cite{RN561,RN555,RN10}.
In summary, Alloy A1 benefits from a synergistic combination of large transformation volume change, dense Ti$_2$Ni precipitates with strained boundaries. 
These features collectively contribute to its large latent heat, reduced hysteresis, and superior mechanical work output at elevated temperatures.

\section{Conclusion}

We have introduced a generative inversion framework for the inverse design of shape memory alloys (SMAs), enabling property-targeted generation of alloy compositions and processing conditions. 
By coupling a pretrained generative adversarial network (GAN) with a differentiable property predictor, the framework performs gradient-based optimization in latent space to identify alloy designs that meet user-defined objectives, such as elevated transformation temperature and high mechanical work output.

The effectiveness of the approach was demonstrated through both computational evaluation and experimental validation. 
The generator successfully captured the complex, multimodal distribution of the SMA design space, while the surrogate model reliably predicted key thermo-mechanical properties.
Latent space optimization enabled rapid and accurate convergence to specified property targets, producing candidate alloys with high fidelity and diversity.
Among five synthesized designs, Ni$_{49.8}$Ti$_{26.4}$Hf$_{18.6}$Zr$_{5.2}$ achieved a martensite start temperature of 404 $^\circ$C and a mechanical work output of 9.9 J/cm$^3$, a transformation enthalpy of 43 J/g,  and a thermal hysteresis of 29 °C, representing a new benchmark in high-temperature SMA performance.
Crystallographic and microstructural analyzes revealed that large transformation volume change, finely dispersed Ti$_2$Ni-type precipitates, and semi-coherent interfaces with localized strain fields together enable large latent heat, reduced hysteresis, and superior mechanical work output at elevated temperatures.

Beyond SMAs, the proposed generative inversion strategy offers a broadly applicable framework for inverse design in complex materials systems,  including high-entropy alloys and functional ceramics.
By enabling targeted exploration of high-dimensional design spaces, this approach provides a powerful tool for accelerating the discovery of advanced materials.

\section*{Acknowledgements}
The authors gratefully acknowledge the support of National Key Research and Development Program of China (2021YFB3802100), National Natural Science Foundation of China (Nos. 52173228, 52271190, and 524B2165), Innovation Capability Support Program of Shaanxi (2024ZG-GCZX-01(1)-06) and Natural Science Foundation Project of Shaanxi Province (Grant No. 2022JM-205).

\section*{Code and data availability}
The code and data are available at https://github.com/mil-licheng/Generative-Inversion-for-Property-Targeted-Materials-Design-Application-to-Shape-Memory-Alloys.

\bibliography{reference}
\end {document}